\documentclass[3p,times]{elsarticle}

\usepackage{xspace}
\usepackage[T1]{fontenc}
\usepackage[scaled]{beramono}
\usepackage{listings}
\usepackage{tikz}
\usepackage{dingbat}
\usepackage{amssymb}
\usepackage{amsmath}
\usepackage{bbding}
\usepackage{pbox}
\usepackage{scrextend}
\usepackage{soul}
\usepackage{multirow,tabularx}

\usetikzlibrary{arrows,positioning,shapes.geometric}

\newcommand{\pycosmo}{\texttt{PyCosmo}\xspace}

\newcommand{\python}{\texttt{Python}\xspace}
\newcommand{\class}{\texttt{CLASS}\xspace}
\newcommand{\camb}{\texttt{CAMB}\xspace}
\newcommand{\cosmics}{\texttt{COSMICS}\xspace}

\begin{document}

\begin{frontmatter}

\title{PyCosmo: An Integrated Cosmological Boltzmann Solver}

\author{Alexandre Refregier$^1$}
\author{Lukas Gamper$^1$}
\author{Adam Amara$^1$}
\author{Lavinia Heisenberg$^2$}
\address{$^1$Department of Physics, ETH Zurich, Wolfgang Pauli Strasse 27, 8093 Zurich, Switzerland}
\address{$^2$Institute for Theoretical Studies, ETH Zurich, Clausiusstrasse 47, 8092 Zurich, Switzerland}

\begin{abstract}

As wide-field surveys yield ever more precise measurements, cosmology has entered a phase of high precision requiring
highly accurate and fast theoretical predictions. At the heart of most cosmological model predictions is a numerical solution of the Einstein-Boltzmann equations governing the evolution of linear perturbations in the Universe. We present \pycosmo, a new Python-based framework to solve this set of equations using a special purpose solver based on symbolic manipulations, automatic generation of C++ code and sparsity optimisation. The code uses a consistency relation of the field equations  to adapt the time step and does not rely on physical approximations for speed-up. After reviewing the system of first-order linear homogeneous differential equations to be solved, we describe the numerical scheme implemented in \pycosmo. We then compare the predictions and performance of the code for the computation of the transfer functions of cosmological perturbations and compare it to existing cosmological Boltzmann codes. We find that we achieve comparable execution times for comparable accuracies. While \pycosmo does not yet have all the features of other codes, our approach is complementary to existing cosmological Boltzmann solvers and can be used as an independent test of their numerical solutions. The symbolic representation of the Einstein-Boltzmann equation system in \pycosmo provides a convenient interface for implementing extended cosmological models.  We also discuss how the \pycosmo framework can also be used as a general framework to compute cosmological quantities as well as observables for both interactive and high-performance batch jobs applications. Information about the \pycosmo package and future code releases are available at 
{\tt http://www.cosmology.ethz.ch/research/software-lab.html}.
\end{abstract}


\end{frontmatter}

\section{Introduction}
\label{sec:introduction}

In order to address the fundamental questions raised by the nature of Dark Matter, Dark Energy and large scale gravity, a number of cosmological experiments are currently underway or in the planning (eg. 
DES\footnote{\tt http://www.darkenergysurvey.org}, DESI\footnote{\tt http://desi.lbl.gov}, LSST\footnote{\tt http://www.lsst.org}, Euclid\footnote{\tt http://sci.esa.int/euclid/}, WFIRST\footnote{\tt http://wfirst.gsfc.nasa.gov}). These experiments aim to achieve the high precision required to tackle these questions and will thus require highly accurate predictions of observables for a wide set of cosmological models.  At the heart of most cosmological model prediction is a numerical solution of the Einstein-Boltzmann equations (see \cite{MB95} and reference therein) governing the linear evolution of perturbations in the Universe. Several codes have thus been developed to produce fast and accurate solutions to this set of first-order linear homogeneous differential equations, such as \cosmics \cite{cosmics}, {\tt CMBFAST} \cite{cmbfast}, {\tt CMBEASY} \cite{cmbeasy}, \camb \cite{camb00}, \class \cite{classI}, with only the latter two being currently maintained. The predictions of these codes are then compared to measurements from cosmological surveys to derive constrains on the parameters of the 
cosmological model using Monte-Carlo Markov Chain techniques (eg. \cite{cosmomc,cosmohammer}). 
Given the central importance of Boltzmann codes to our current constraints on the cosmological model and the well known numerical difficulty to solve this set of equations (see e.g. \cite{Nadkarni2017} for a mathematical analysis, and references therein), it is important to explore 
different numerical schemes for the solutions of the differential equations. 

In this paper, we present \pycosmo, a new Python-based framework to solve the Einstein-Boltzmann equations using a special purpose solver based on symbolic manipulations, automatic generation of C++ code and sparsity optimisation. The code uses a consistency relation of the field equations to adapt the time step and does not rely on physical approximations for speed-up. We study the accuracy and performance of \pycosmo for the computation of the transfer functions of cosmological perturbations in the newtonian gauge in a flat universe, and compare it to existing codes. Our approach is complementary to existing cosmological Boltzmann solvers that are based on more general differential equation solvers and physical approximations, and can be used as an independent test of their numerical solutions. We discuss how the symbolic representation of the Einstein-Boltzmann equation system in \pycosmo provides a convenient interface for theorists to rapidly implement new cosmological models.  We also discuss how the \pycosmo framework can also be used as a general framework to compute cosmological quantities as well as observables for both interactive and for high-performance batch jobs applications, drawing upon the earlier \texttt{IDL} cosmological package \texttt{ICosmo} \cite{icosmo}. 

The paper is organised as follows. In section \ref{sec:eb_eq}, we describe the set of Einstein-Boltzmann equations describing the linear growth of cosmological structures . Section \ref{sec:numerics} describes our implementation scheme for deriving numerical solutions to this
set of equations. In section \ref{sec:results} we study the performance of \pycosmo in terms of numerical precision and speed and compare it to existing codes. Our conclusions are described in section \ref{sec:conclusion}.

\section{Einstein-Boltzmann Equations}
\label{sec:eb_eq}

\subsection{Linear Perturbations}
After solving the evolution of the scale factor $a$ and the Hubble parameter $H$ using the Friedmann equation (see eg. \cite{dod03}), we can consider the linear evolution of scalar perturbations. For this purpose, we choose the newtonian gauge
in a flat $\Lambda$CDM cosmology.  The evolution of perturbations are thus governed by the Einstein-Boltzmann equations which, in this case and to linear order, are given by (\cite{MB95} with the conventions of \cite{dod03})
\begin{eqnarray}
\label{eq:eb_first}
\dot{\delta}  & = & - ku -3\dot{\Phi}  \label{eq:continuity} \label{eq:eb_nophoton}\\
\dot{u} & = & - \frac{\dot{a}}{a}u + k \Psi\\
\dot{\delta_b} & = & -ku_b -3\dot{\Phi} \\
\dot{u_b} & = & -\frac{\dot{a}}{a}u_b + k \Psi + k c_s^2 \delta_b + \frac{\dot{\tau}}{R} [ u_b-3\Theta_1]  \\
\dot{\Theta}_0 & = & -k \Theta_1 - \dot{\Phi} \\
\dot{\Theta}_1 & = & \frac{k}{3} \left[ \Theta_0-2\Theta_2+ \Psi \right] + \dot{\tau} \left[ \Theta_1 - \frac{u_b}{3} \right]  \\
\dot{\Theta}_2 & = & \frac{k}{5} \left[ 2 \Theta_1 - 3 \Theta_3 \right] + \dot{\tau} \left[ \Theta_2 - \frac{\Pi}{10} \right] \\
\dot{\Theta}_l & = & \frac{k}{2 l +1} \left[ l \Theta_{l-1} - (l+1) \Theta_{l+1} \right] + \dot{\tau} \Theta_l,~~l>2 \label{eq:theta_l} \\
\dot{\Theta}_{Pl} & = & \frac{k}{2 l +1} \left[ l \Theta_{P(l-1)} - (l+1) \Theta_{P(l+1)} \right] + \dot{\tau} \left[ \Theta_{Pl}-\frac{\Pi}{2} (\delta_{l,0}+\frac{\delta_{l,2}}{5}) \right] \label{eq:theta_pl} \\
\dot{\mathcal N}_0 & = & -k \mathcal N_1 - \dot{\Phi}  \label{eq:n_l_first}\\
\dot{\mathcal N}_1 & = & \frac{k}{3} \left[ \mathcal N_0-2\mathcal N_2+ \Psi \right]  \\
\dot{\mathcal N}_l & = & \frac{k}{2 l +1} \left[ l \mathcal N_{l-1} - (l+1) \mathcal N_{l+1} \right],~~l>1 \label{eq:n_l} \\
k^{2}\Phi+3 \frac{\dot{a}}{a} \left( \dot{\Phi}- \frac{\dot{a}}{a} \Psi \right) & = & 4 \pi G a^2 \left[ \rho_m \delta_m + 4 \rho_r \Theta_{r0} \right] \label{eq:eb_phi_tt},
\label{eq:eb_last}
\end{eqnarray}
where  dot denotes derivatives with respect to conformal time $\eta$, $\delta$ ($\delta_b$) and $u$ ($u_b$) are the density and velocity perturbations for the dark matter (baryons), $\Theta_l$ and $\Theta_{Pl}$ are the photon temperature and polarization multipole moments, ${\mathcal N}_l$ are the multipole moments of the (massless) neutrino temperature, and $\Pi = \Theta_2+\Theta_{P0}+\Theta_{P2}$. The baryon-to-photon ratio is given by $R = 3\rho_b/(4\rho_{\gamma})$,  $\tau$ is the Thomson scattering optical depth  and $c_s$ is the baryon sound speed. The subscripts $r$ and $m$ refer to the density-weighted sum of all radiation and matter components, and $\rho_r$ and $\rho_m$ is the mean density in each of these components.

The gravitational potential fields $\Phi$ and $\Psi$ describe scalar perturbations to the metric
$ds^{2} = - (1+2 \Psi) dt^{2} + a^{2}  ( 1+2 \Phi) d\vec{x}^2$, and are related by the longitudinal traceless space-space parts of the Einstein equation by means of the algebraic equation
\begin{equation}
\label{eq:phi_psi}
k^2 (\Phi+\Psi) =  -32 \pi G a^2 \rho_r \Theta_{r2}.
\end{equation}
Note that an alternative to the time-time Einstein equation (Eq.~\ref{eq:eb_phi_tt}) is its time-space component given by
\begin{equation}
\dot{\Phi}- \frac{\dot{a}}{a} \Psi = - 4\pi G \frac{a^2}{k} \left[ \rho_m u_m + 4 \rho_r \Theta_{r1} \right].
\label{eq:eb_phi_alter}
\end{equation}

\subsection{Initial Conditions}
\label{initial}
We assume initial conditions arising from inflation for which the primordial power spectrum of the potential $\Phi$ 
follows
\begin{equation}
\label{eq:p_phi_p}
P_{\Phi}(k) = \frac{50 \pi^2}{9 k^3} \left( \frac{k}{H_0} \right)^{n-1} \delta_{H}^2  \left( \frac{\Omega_m}{D(a=1)} \right)^2,
\end{equation}
where $\delta_H$ is a normalisation parameter, $D(a)$ is the late time growth factor (normalised to $D(a)=a$ in the
matter era), $n$ is the scalar spectral index, and $H_0$ is the present value of the Hubble parameter.
For adiabatic initial conditions, the other fields at early times are given by \cite{MB95}
\begin{eqnarray}
\label{eq:ic}
\Phi & = &- \left( 1 + \frac{2}{5} R_{\nu}  \right) \Psi  \nonumber \\
\delta&=&\delta_b=3\Theta_0=3\mathcal N_0=-\frac{3}{2} \Psi  \\
u & = & u_b=3\Theta_1=3 \mathcal N_1=\frac{1}{2}k\eta \Psi \nonumber \\
\mathcal N_2 & = & \frac{1}{30}  k\eta \Psi,  \nonumber
 \end{eqnarray}
where $R_{\nu}=(\rho_\nu+P_\nu)/(\rho+P)=\Omega_\nu/\left(\frac{3}{4} \Omega_m a+ \Omega_r\right)$  is the neutrino ratio,
and all the other perturbation fields are set to 0 at the initial time.

\subsection{Practicalities}
In practice, the hierarchy of moments for the photons and the neutrinos can be truncated to a maximum multipole $l_{\rm max}$
by replacing Equations~(\ref{eq:theta_l}-\ref{eq:theta_pl}) for $\dot{\Theta}_l$ at $l=l_{\rm max}$ with \cite{MB95}
\begin{equation}
\label{eq:lmax_truncate}
\dot{\Theta}_l \simeq  k \Theta_{l-1} - \frac{l+1}{\eta} \Theta_l + \dot{\tau} \Theta_l,
\end{equation}
and similarly for $\Theta_{Pl}$ and for ${\mathcal N}_l$, but without the Thomson scattering term in the latter case.  

The optical depth $\tau$ and the sound speed $c_s$ can be pre-computed using public recombination codes such as {\tt RECFAST} \cite{1999ApJ...523L...1S,2000ApJS..128..407S}, {\tt RECFAST++} \cite{1999ApJ...523L...1S,2010MNRAS.402.1221C,2010MNRAS.407..599C,2010MNRAS.403..439R,2011MNRAS.412..748C} or {\tt COSMOSPEC} \cite{2016MNRAS.456.3494C}. In \pycosmo, we have implemented an interface to {\tt RECFAST++}, as well as the possibility of external input recombination variables for comparisons with other Boltzmann codes.

To test for numerical accuracy, we use the redundancy of the Einstein field equations following \cite{MB95,1995astro.ph..6070B}. From the algebraic relation in equation (\ref{eq:phi_psi}), we can express the non-dynamical gravitational field $\Psi$ in terms of $\Phi$. From the time-space component of the Einstein equations (Eq.~\ref{eq:eb_phi_alter}), we can express the time derivative of the gravitational field $\Phi$ in terms of the other fields. The expression for $\dot{\Phi}$ can then be plugged back into the time-time component of the Einstein equations (Eq.~\ref{eq:eb_phi_tt}), which results in an algebraic equation for $\Phi$. This reflects the fact, that the gravitational field $\Phi$ is not a dynamical degree of freedom and can be solved algebraically in terms of the remaining relativistic and non-relativistic matter fields. We will use this combination of the two Einstein equations (\ref{eq:eb_phi_tt}) and (\ref{eq:eb_phi_alter}) in order to test our numerical accuracy. We will denote the resulting algebraic equation by the dimensionless parameter defined as \cite{MB95,1995astro.ph..6070B}
\begin{equation}
\label{eq:econ}
\epsilon = \left[ - \frac{2}{3} \left( \frac{k}{a H_0}\right)^2 \Phi + \left( \Omega_m \delta_m a^{-3}  + 4  \Omega_{r} \Theta_{r0} a^{-4} \right)
 + 3 \frac{aH}{k} \left( \Omega_m u_m  a^{-3}  + 4  \Omega_r \Theta_{r1} a^{-4}  \right) \right] \left(  \Omega_m a^{-3}  + \Omega_r a^{-4} + \Omega_\Lambda \right)^{-1}
\end{equation}
which ought to vanish for the consistency of the equations and thus gives a relative measure of numerical errors. We thus use $\epsilon$ to adapt our time steps and choose to use eq.~(\ref{eq:eb_phi_tt}) in our Einstein-Boltzmann system.

The initial conditions given in equation~(\ref{eq:ic}) are implemented using the scheme in \class \cite{classI} which includes higher order terms for improved numerical precisions at early time. We also choose the initial time of the initial condition
as in \cosmics, i.e. an initial conformal time $\eta= \min [ 10^{-3} k^{-1} , 10^{-1} h^{-1} {\rm Mpc} ]$.

\section{Numerical solution scheme}
\label{sec:numerics}

In this section, we describe our method to solve the Einstein-Boltzmann Equations
(Eqs.~\ref{eq:eb_first}-\ref{eq:eb_last}) to derive the transfer functions for the different perturbation fields.

\subsection{Differential equation system}

For each wavenumber $k$, we solve the linear homogeneous non-autonomous system of $8+3 l_{\rm max}$ coupled differential  equations (Eqs.~\ref{eq:eb_nophoton}-\ref{eq:eb_last}, with the algebraic equation~\ref{eq:phi_psi}) which can be recast in the form 
\begin{equation}
\label{eq:eb_mform}
\dot{\mathbf y}(k,\eta) = {\mathbf J}(k,\eta)  {\mathbf y}(k,\eta)
\end{equation}
where ${\mathbf y}(k,\eta)=(\Phi, \delta,u, \delta_b,u_b,\Theta_0,\Theta_{P0}, \mathcal N_0,\Theta_1,\Theta_{P1},\mathcal N_1,\ldots,\Theta_{l_{\rm max}},\Theta_{Pl_{\rm max}}, \mathcal N_{l_{\rm max}})$ is the vector of the $8+3 l_{\rm max}$ perturbation fields, dot denotes, as earlier, a
derivative in terms of the conformal time $\eta$, and ${\mathbf J}(k,\eta)$ is a $k$- and time-dependent Jacobian matrix. 
This was easily achievable by solving for $\dot{\Phi}$ in Equation~\ref{eq:eb_last} and substituting it in the equations which have
an explicite dependence on it. For convenience, we perform a change of variable to change the time variable $\eta$ to $\ln a$ for the purpose of the integration, which has the advantage of reducing the dynamic range of the time steps during the integration in practice.

\subsection{Solver}

To solve the set of Einstein-Boltzmann equations, we have implemented a novel solver scheme based on a combination of symbolic
manipulations, automatic {\tt C++} code generation and numerical integration tuned to this problem. 

One of the central components in this scheme is to represent  the set of linear differential equations of the perturbations (Eqs.~\ref{eq:eb_nophoton}-\ref{eq:eb_last}) in symbolic form. We implement this using the {\tt sympy} package. This has the advantage of allowing the package to symbolically manipulate the equations within the \python framework. Moreover, starting from a symbolic representation provides a convenient interface for theorists who wish to modify the set of equations to be integrated. This also allows the code to generate symbolic equations automatically for each of the multipole moments $l > 2$ for the photons and neutrinos up to a set maximum multipole moment $l_{\rm max}$ using their recurrence relations. Furthermore, the jacobian matrix ${\mathbf J}$ (Eq.~\ref{eq:eb_mform}) is  generated in symbolic form by applying the {\tt jacobian} function of {\tt sympy} to the set of differential equations. The symbolic form is also used to
change the time variable from $\eta$ to $\ln a$ as discussed above.

As the numerical integration scheme we use the second order Backward-Differentiation-Formula method (BDF2) with variable time-step (see e.g. \cite{Eckert2004}). This integration scheme has the advantage of being implicit, multi-step, second order and A-stable
unlike multi-step methods of higher-order. It can be written as
\begin{equation}
\frac{1+2\tau_n}{1+\tau_n} {\mathbf y}_{n+1} - (1+\tau_n) {\mathbf y}_{n}+\frac{\tau_n^2}{1+\tau_n}{\mathbf y}_{n-1} = \Delta t_{n} {\mathbf J} {\mathbf y}_{n+1}, 
\end{equation}
where ${\mathbf y}_{n}={\mathbf y}(t_n)$, $\Delta t_{n}=t_{n+1}-t_{n}$ is the variable time step, $\tau_n=\Delta t_n/\Delta t_{n-1}$, and the time variable is $t=\ln a$ in our implementation. We recast this equation into the form
\begin{equation}
{\mathbf A} {\mathbf y}_{n+1} = {\mathbf b}({\mathbf y}_{n},{\mathbf y}_{n-1}),
\end{equation}
where ${\mathbf A} = (1+2\tau_n)(1+\tau_n)^{-1} {\mathbb I} - \Delta t_{n} {\mathbf J}$ and ${\mathbf b}= (1+\tau_n) {\mathbf y}_{n}-\tau_n^2(1+\tau_n)^{-1}{\mathbf y}_{n-1}$.
We symbolically generate the matrix ${\mathbf A}$ and simplify it using the {\tt simplify} function in {\tt sympy}.

To solve this equation for $y_{n+1}$ in terms of $y_{n-1}$ and $y_{n}$, we use LU factorisation with Partial Pivoting  followed by a triangle-solve algorithm (see e.g. \cite{Menon2004} for a description).  We first do this symbolically assuming that no partial pivoting is necessary in the LU factorisation. In particular, we evaluate the $A$ matrix by isolating common sub-expressions using the {\tt cse} function in {\tt sympy}.

Another central feature of our method is to automatically generate optimised {\tt C++} code, an approach that
uses heritage from our just-in-time \python compiler {\tt HOPE} \cite{hope}. This is done by
transforming the symbolic {\tt sympy} expressions into {\tt C++} code using a visitor pattern. In the process,
the {\tt C++} is optimised by replacing integer powers by multiple products. The {\tt C++} code is then generated to perform the LU factorisation considering only the non-zero elements in the ${\mathbf A}$ matrix and by unrolling the loops. This improves performance significantly as the ${\mathbf A}$  matrix has  $(8+3 l_{\rm max})^2$  elements but is very sparse with only of order $8l_{\rm max}$ non-zero elements (see recurrence relations for the photons and neutrino moments) (Eqs.~\ref{eq:theta_l} \&\ref{eq:n_l}). We then generate the solution for $y_{n+1}$ using the triangle-solve algorithm. We use the {\tt Python} {\tt distutils} package to compile the {\tt C++} code into a dynamic library which we import into {\tt Python}.  

\begin{figure}[t]
\begin{center}
\includegraphics[width=1.0\linewidth]{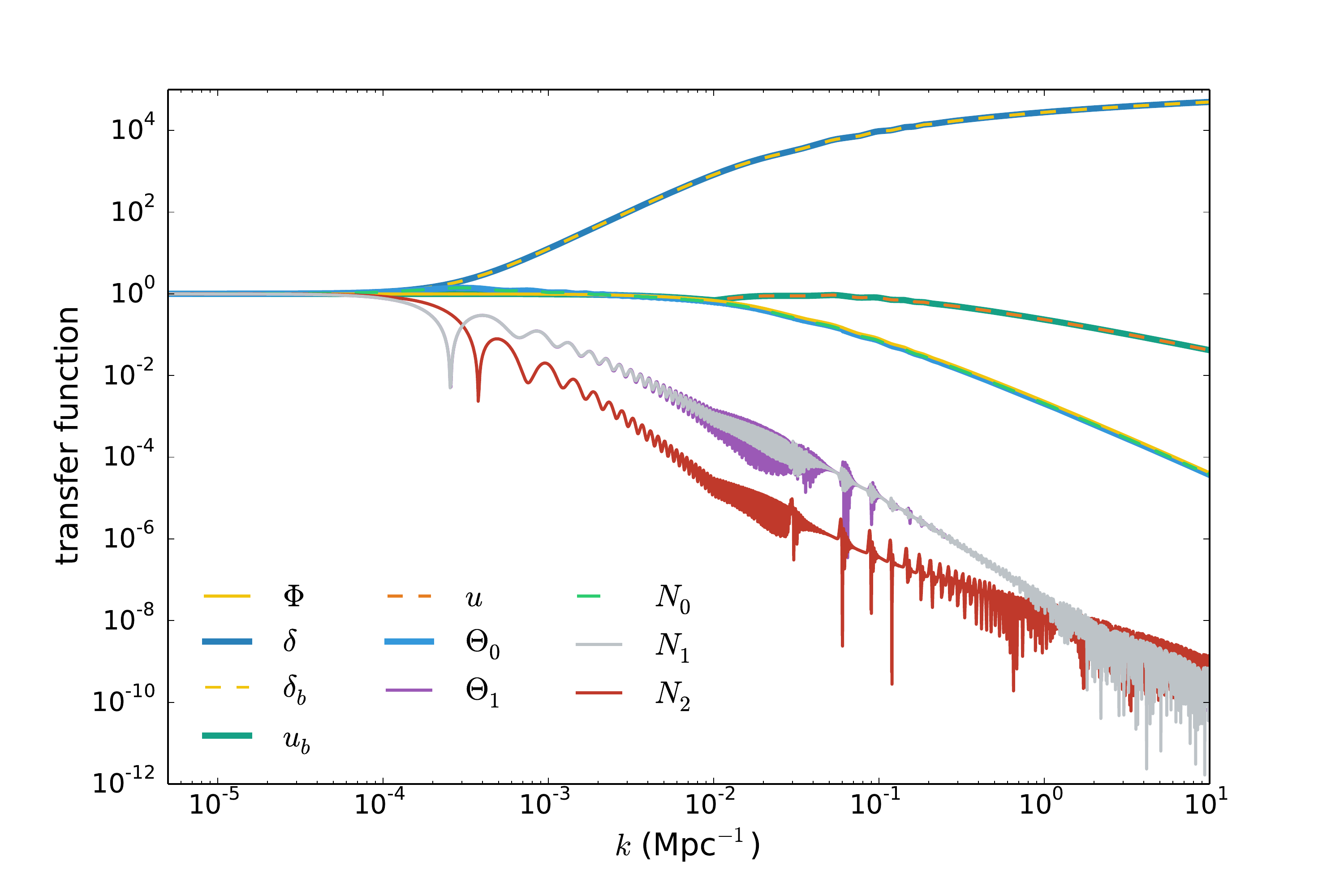}
\end{center}
\caption{Transfer function T(k)  for the main perturbation fields at $a=1$ as computed by \pycosmo. They were normalised such that $T(k) \rightarrow 1$ as $k \rightarrow 0$ and the reference accuracy setting for \pycosmo were used (see text).
} 
\label{fig:t_k}
\end{figure}

During runtime, the algorithm checks whether partial pivoting is necessary. If this is the case, it builds a graph to track the additional row permutations that are needed. New {\tt C++} code can then be generated for each new permutation optimised as described above. For speed up, we implemented a caching scheme for the symbolic BDF equations. In practice, only a few dozen permutations are needed.

For the adaptive control of the time step $\Delta t$, we impose the following constraints: 
\begin{itemize}
\item The parameter $\epsilon$ (Eq.~\ref{eq:econ}) that we derived from the consistency of the field equations is used as a dimensionless measure of numerical errors. We impose that
it remains within a set upper and lower limit after smoothing on a set time scale $\Delta t_{\rm smooth}$ typically set to 0.1.
If the upper limit is exceeded, we step back by $\Delta t_{\rm smooth}$ and reduce the timestep $\Delta t$ by 50\% at that point. If $\epsilon$ becomes smaller than the lower limit, we increase the timestep by 25\%. For safety, we do not allow the time step to be further modified for another smoothing time scale.
\item We also impose that the time step $\Delta t$ is smaller than the expansion time $\eta a H$ and the Courant time $aH/k$ (for $t=\ln a$) as in \cosmics. If these limits are exceed we reduce the time scale by 50\% using the smoothing time scale described above. 

\end{itemize}

\section{Results}
\label{sec:results}

\begin{figure}[t]
\begin{center}
\includegraphics[width=1.0\linewidth]{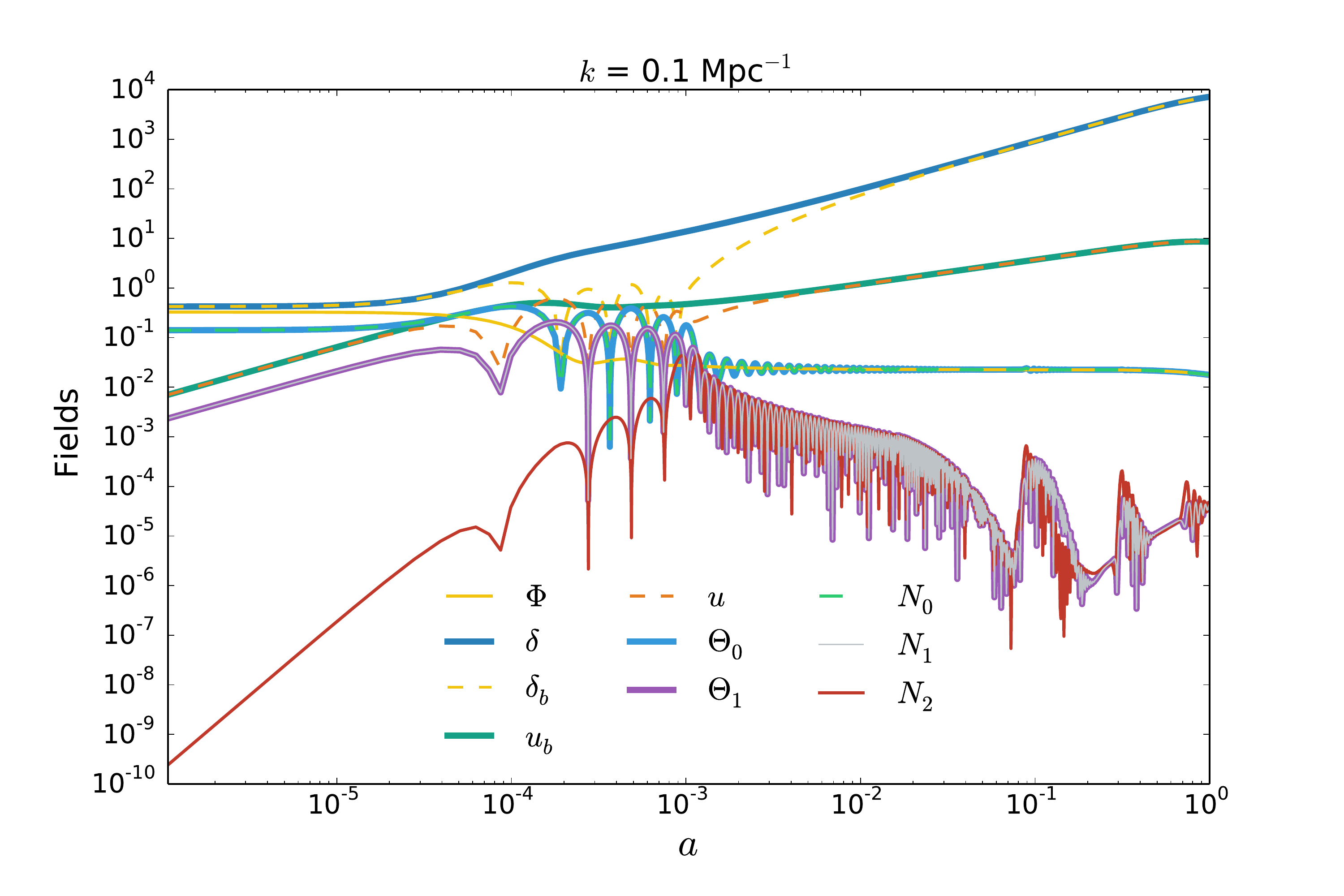}
\end{center}
\caption{Main perturbation fields as a function of the scale factor $a$ computed by \pycosmo with the reference accuracy
setting and for the wavenumber $k=0.1$ Mpc$^{-1}$.} 
\label{fig:t_a}
\end{figure}

After presenting the implementation of our solver, we now summarise our results on its performance for the transfer
functions $T(k)$ of the perturbations. Figure~\ref{fig:t_k} shows the transfer functions as function of the wavenumber $k$ for  the main perturbation fields at $a=1$ as computed by \pycosmo. For this purpose and below, we assumed a flat $\Lambda$CDM cosmology with $\Omega_m=0.3$, $\Omega_\Lambda=0.7$, $\Omega_b=0.06$ and $h=0.7$. The transfer functions were normalised such that $T(k) \rightarrow 1$ as $k \rightarrow 0$. For this figure, we have used a high accuracy \pycosmo setting of $l_{\rm max}=200$ and $\epsilon<10^{-6}$ which we use as a reference setting.

Figure~\ref{fig:t_a} illustrates the time evolution of the perturbations by showing the main perturbation fields as a function of the scale factor $a$ computed by \pycosmo with the same setting. The normalisation was chosen to match that of $\Phi$ in the earlier figure and the wave number was chosen to be $k=0.1$ Mpc$^{-1}$.

\begin{figure}[t]
\begin{center}
\includegraphics[width=1.0\linewidth]{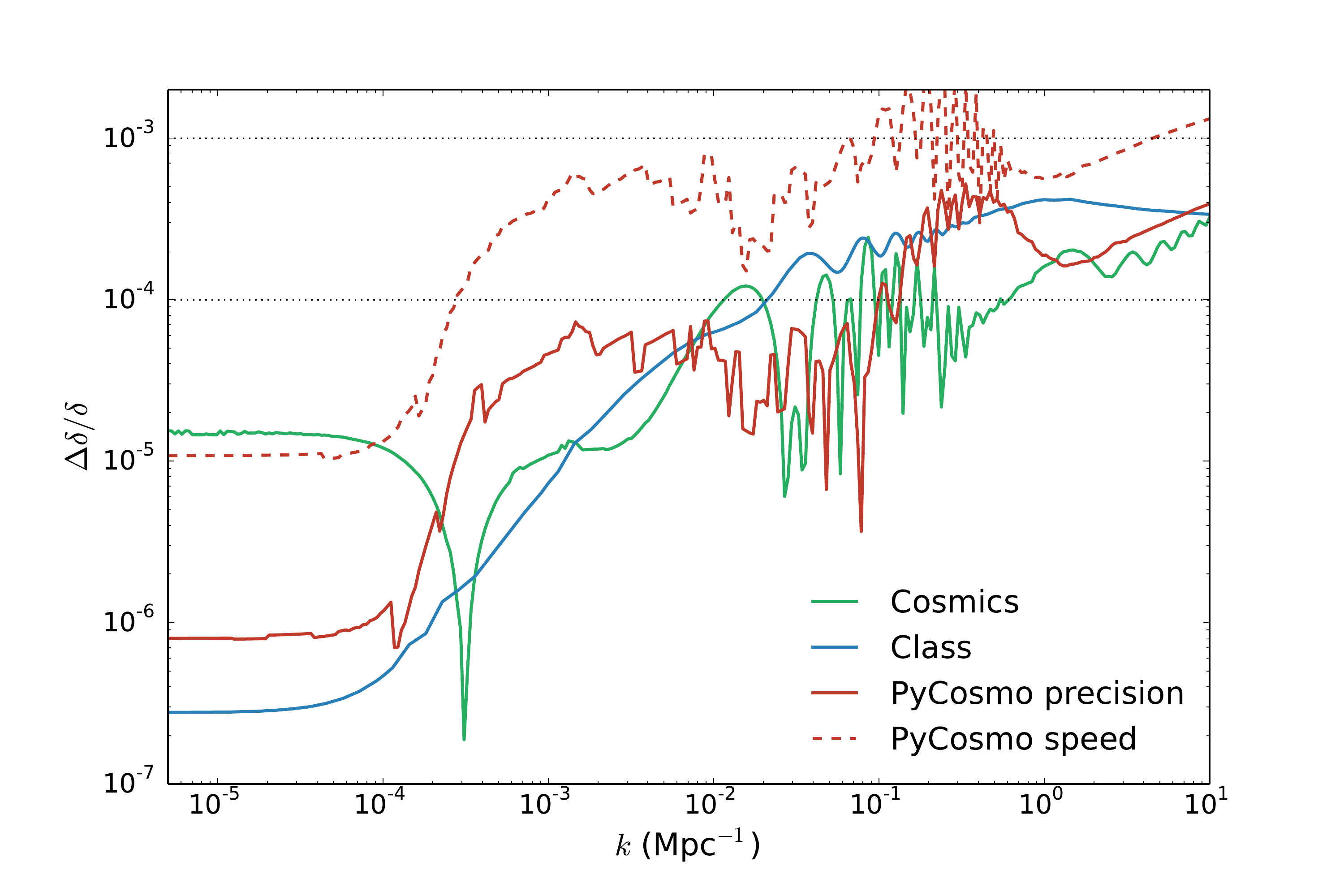}
\end{center}
\caption{Relative difference dark matter density $\delta$ at $a=1$ between the different Boltzmann codes. For \pycosmo different accuracy settings are shown (see text).} 
\label{fig:diff_k}
\end{figure}

Figure~\ref{fig:diff_k} shows the relative difference in the dark matter density perturbation field $\delta$ in the newtonian gauge between the Boltzmann codes \pycosmo, \class and \cosmics, as a function of wavenumber $k$ at $a=1$.  All results are shown
relative to \pycosmo with the reference setting given above. \class was run using the high accuracy precision file 
{\tt pk\_ref.pre} available in the public distribution version 2.5.0, while \cosmics was run using its default settings. 
For \pycosmo, we consider a setting chosen for precision with $l_{\rm max}=50$ and $\epsilon<3\times10^{-5}$, and a setting chosen for speed with $l_{\rm max}=30$ and $\epsilon<3\times10^{-4}$. Also in this and the other comparisons below, the recombination variables $c_s$ and $\tau$ for \pycosmo were taken to be that of \class to focus the comparison on the Boltzmann solvers rather than the recombination codes. Since \camb does not have a newtonian gauge setting, we do not include this code in this comparison, but it was shown to agree with \class at the 0.01\% level for the matter transfer function with this high precision setting \cite{classIII}.

As the figure shows, we find that \class, \cosmics and \pycosmo with the precision setting achieve an accuracy better
than about 0.01\% for $k<10^{-2}$ Mpc$^{-1}$ and 0.05\% for $k<10$ Mpc$^{-1}$ compared to the reference run. 
These differences are comparable  to those found between \class and \camb at high accuracy settings by \cite{classIII}.
When used in the speed setting, the figure shows that \pycosmo achieves an accuracy of about 0.1\% for $k<10$ Mpc$^{-1}$.
 
\begin{table}
\centering
\caption{Execution times (sec) from the best of  5 executions on a single core of a MacBook Pro with 2.7 GHz intel Core i7 and 16GB of RAM for \pycosmo and \class with two different accuracy settings.
\label{tab:times}}
\begin{tabular}{l|lll|lll}
\hline 
  &   \multicolumn{3}{c}{Precision}  & \multicolumn{3}{|c}{Speed}  \\ 
$k_{\rm max}$ &  &  \pycosmo & \class & &  \pycosmo & \class \\
(Mpc$^{-1}$)& $N_k$ & $l_{\rm max}=50$, $\epsilon<3\times10^{-5}$ & {\tt pk\_ref.pre} & $N_k$ & $l_{\rm max}=30$, $\epsilon<3\times10^{-4}$ &  {\tt cl\_permille.pre} \\
\hline
0.1 & 72 & 1.2 & 1.1 & 55 & 0.28 & 0.20 \\
1.0 & 129 & 2.5 & 2.8 & 112 & 0.68 & 0.42\\
10 & 139 & 3.1 & 4.9 & 122 & 1.0 & 0.67\\
29 & 144 & 4.2 & 9.5 & 127 & 1.6 & 1.3\\
\hline 
\end{tabular}
\end{table}

The performance of the different codes can be compared by considering their execution time for different
accuracy settings. For this purpose, we consider the timings derived from the best of  5 executions on a single  core of a MacBook Pro with 2.7 GHz intel Core i7 and 16GB or RAM. Table~\ref{tab:times}, describes the resulting
execution times for \pycosmo with the precision and speed settings described above. The timings for \class are
given for a precision setting as described above and a speed setting obtained using the {\tt cl\_permille.pre} precision file.
The settings were chosen to achieve an accuracies of about 0.01\% and better than 0.1\%, respectively, for the
transfer function \cite{classIII},  which is comparable to the accuracy of \pycosmo for the corresponding 
settings. The table also shows results for different maximal wavenumber $k_{\rm max}$ and number, $N_{k}$,  of $k$ values evaluated. As the table shows, the timings for \pycosmo and \class are comparable for both
settings and different $k_{\rm max}$.

\section{Conclusions}
\label{sec:conclusion}

In this paper, we presented \pycosmo, a new cosmological Boltzmann code, and described its special purpose solver based on symbolic manipulations, automatic generation of C++ code and sparsity optimisation. We also showed how the code does not rely on physical approximations for speed-up and uses the consistency among the field equations to adapt the time step. 
The deviations of the fields from this algebraic relation gives us a self-consistent control of the numerical error. From a performance test, we find that the code achieve comparable execution times for comparable accuracies to current Boltzmann solvers for the computation of the transfer functions in the newtonian gauge. \pycosmo currently does not have all the features of these other solvers, but it is complementary to existing cosmological Boltzmann solvers which rely on more general differential equation solvers and physical approximations, and can be used as an independent test of their numerical solutions. The symbolic representation of the Einstein-Boltzmann equation system in \pycosmo provides a convenient interface to implement extended cosmological models. In future works, we aim to implement more general models beyond flat $\Lambda$CDM and with more general gravitational field theories beyond General Relativity. The \pycosmo framework can also be used as a general framework to compute cosmological quantities as well as observables for both interactive and for high-performance batch jobs applications. The computation with \pycosmo of observables such as the CMB angular power spectrum, weak lensing statistics and galaxy clustering statistics is left to a future publication. Information about the \pycosmo package and future code releases are available at {\tt http://www.cosmology.ethz.ch/research/software-lab.html}.

 \section*{Acknowledgements}
The authors thank Andrina Nicola, Aseem Paranjape, Joel Akeret, Oliver Hahn, Sharvari Narkarni-Ghosh and Caroline Bertemes  for useful discussions , and Julien Lesgourgues for useful comments and providing \class high precision settings for the newtonian gauge. LH thanks financial support from Dr. Max R\"{o}sller, the Walter Haefner Foundation and the ETH Zurich Foundation.

\bibliographystyle{elsarticle-num}

\bibliography{pycosmo}

\end{document}